\documentclass[twocolumn,preprintnumbers,superscriptaddress,nofootinbib,aps,prd,floatfix]{revtex4}
\pdfoutput=1

\usepackage{amsmath,amssymb}
\usepackage{wasysym}
\usepackage{graphicx}
\usepackage{color,array,subfigure,slashed}
\usepackage[scientific-notation=true]{siunitx}
\usepackage[dvipsnames]{xcolor}
\usepackage{ulem}

\newcommand{\be}{\begin{eqnarray}}
\newcommand{\ee}{\end{eqnarray}}

\newcommand{\bee}{\begin{eqnarray}}
\newcommand{\eee}{\end{eqnarray}}
\newcommand{\beeq}{\begin{equation}}
\newcommand{\eeeq}{\end{equation}}
\newcommand{\gev}{{\text{GeV}}}

\numberwithin{equation}{section}

\begin{document}

\title{Perturbative Higgs CP violation, unitarity and phenomenology}

\begin{abstract}
Perturbative probability conservation provides a strong constraint on the presence of new interactions of the Higgs boson. In this work we consider CP violating Higgs interactions in conjunction with unitarity constraints in the gauge-Higgs and fermion-Higgs sectors. Injecting signal strength measurements of the recently discovered Higgs boson allows us to make concrete and correlated predictions of how CP-violation in the Higgs sector can be directly constrained through collider searches for either characteristic new states or tell-tale enhancements in multi-Higgs processes.
\end{abstract}

\author{Christoph Englert} \email{christoph.englert@glasgow.ac.uk}
\affiliation{SUPA, School of Physics and Astronomy, University of
  Glasgow,\\Glasgow G12 8QQ, UK \\[0.1cm]}

\author{Karl Nordstr\"om} \email{k.nordstrom.1@research.gla.ac.uk}
\affiliation{SUPA, School of Physics and Astronomy, University of
  Glasgow,\\Glasgow G12 8QQ, UK \\[0.1cm]}

\author{Kazuki Sakurai} \email{kazuki.sakurai@durham.ac.uk}
\affiliation{Institute for Particle Physics Phenomenology, Department
  of Physics,\\Durham University, Durham DH1 3LE, UK\\[0.1cm]}
\affiliation{Institute of Theoretical Physics, Faculty of Physics,\\University of Warsaw, ul.~Pasteura 5, PL--02--093 Warsaw, Poland\\[0.1cm]}

\author{Michael Spannowsky} \email{michael.spannowsky@durham.ac.uk}
\affiliation{Institute for Particle Physics Phenomenology, Department
  of Physics,\\Durham University, Durham DH1 3LE, UK\\[0.1cm]}

\preprint{IPPP/16/109, DCPT/16/218}

\maketitle


\section{Introduction}
\label{sec:intro}
There is strong evidence that the Higgs boson which was discovered in 2012 can be characterised by a dominant CP-even coupling pattern to gauge bosons~\cite{Chatrchyan:2012jja,Aad:2015mxa}. The sensitivity of this measurement is driven by large modified production rates compared to the Standard Model (SM) if CP-odd couplings were dominant~\cite{Gao:2010qx,Bolognesi:2012mm,Englert:2012xt,Englert:2012ct,Corbett:2012dm,Ellis:2013ywa}, as well as different kinematics if cross section information is not included in the analysis~\cite{Hankele:2006ma,Chatrchyan:2012jja,Ellis:2013yxa, Harnik:2013aja,Bishara:2013vya,Dolan:2014upa,Aad:2015mxa,Khachatryan:2014kca,Khachatryan:2016tnr,CMS:2016ilx}.  

The observation of Higgs boson decays to electroweak bosons $h\to ZZ,WW$~\cite{Aad:2015lha,Khachatryan:2015yvw,Aad:2016lvc,Khachatryan:2016vnn} is already a strong indication of a CP-even character of the gauge-Higgs interactions. A CP-odd interaction parameterised by ${\cal{L}} \supset g_{Z\tilde Z h}/v \, h Z^{\mu\nu}\tilde{Z}_{\mu\nu}$\footnote{We understand $h$ fluctuation of the Higgs doublet field $H$ around the vacuum expectation value $v= \sqrt{2} \langle H \rangle\simeq 246~\gev$.} which overpowers the ${\cal{L}}\supset g_{hhZ}\, m_Z \, h Z^{\mu}Z_{\mu}$ term that follows from gauge boson mass generation through electroweak symmetry breaking (EWSB) would imply the breakdown of otherwise successful perturbation theory, only avoided if the longitudinal gauge boson degrees of freedom are generated by a mechanism which is not directly related to the observed Higgs boson with $m_h\simeq 125~\gev$.
Taking the measurements in the $ZZ$ channel at face value, the latter would need to be accompanied by a low scale of perturbative unitarity violation, well below the TeV scale, which is typically mended by either resolving a potential substructure responsible for the TeV scale or by accessing new resonant degrees of freedom. However, the Large Hadron Collider (LHC) has already explored regions well beyond this regime without any evidence of neither weakly nor strongly-coupled degrees of freedom. In this sense, the statistically significant observation of $pp\to h\to ZZ$ alone does cement the very character of mostly CP-even couplings to vector bosons, which is a generic property of spontaneous symmetry breaking directly linked with perturbative unitarity of the Higgs-gauge sector~\cite{Cornwall:1973tb,Cornwall:1974km}.

Strong constraints on the CP violating interactions are typically inferred from flavor and electric dipole measurements~\cite{Regan:2002ta,Baker:2006ts,Griffith:2009zz,Brod:2013cka,Baron:2013eja,Chien:2015xha,Cirigliano:2016njn}. These indirect probes of CP violation (which in the EDM context are strongest for interactions with first or second generation fermions) need to be contrasted with direct searches as performed by ATLAS and CMS. It is, therefore, natural to ask how CP-violation can be accommodated by current Higgs measurements, in particular by the recent combination of ATLAS and CMS data~\cite{Khachatryan:2016vau}.
Given the absence of any conclusive hints for new resonant physics around the TeV scale, and taking into account the aforementioned unitarity-related issues, we can expect that a low energy effective formulation of TeV scale physics will reflect the imprint of a ``good'' probabilistic behavior of the underlying UV model. Understanding an effective theory formulation as the tool of mediating measurements between theories with widely separated scales, large fundamental CP-violating effects at a scale that lies well above the electroweak scale could therefore present themselves at low scales in the guise of operators that do not immediately imply unitarity violation close to the TeV scale. Another possibility is the presence of additional intermediate degrees of freedom which could mend whatever unitarity violation that seems to be present above the TeV scale. Put differently, if no new particles are present, unitarity imposes a well-defined bias on the perturbative expansion of new physics effects in terms of a dimension six extended SM effective field theory framework~\cite{Buchmuller:1985jz,Hagiwara:1986vm,Grzadkowski:2010es}
\begin{equation}
	\label{eq:dim6l}
	{\cal{L}}={\cal{L}}_{\text{SM}} + \sum_i {{C_i}(\mu^2)\over \Lambda_i^2} {\cal{O}}^i\,.
\end{equation}
This hierarchy will be fully reflected by the Wilson coefficients if we choose all $\Lambda_i\equiv\Lambda\gg v$ in Eq.~\eqref{eq:dim6l} and limit ourselves to weakly-coupled UV theories. The latter point is required to give perturbative unitarity violation a well-defined meaning.

In this paper we analyse the tree-level interplay of CP-violation in the fermion-Higgs and gauge-Higgs sectors and unitarity using the tools of effective field theory. Assuming that amplitudes are well-behaved to high energies, we identify operators in Sec.~\ref{sec:eff}, which are largely unconstrained by tree-level unitarity requirements. Using recent signal strength measurements as reported by ATLAS and CMS in Ref.~\cite{Khachatryan:2016vau}, we analyse the direct phenomenological implications of allowed CP violation in the Higgs sector for future LHC exotics searches in Sec.~\ref{sec:coll}. We provide a summary of this work and offer conclusions in Sec.~\ref{sec:conclusions}.

\section{Unitarity and CP violating Operators}
\label{sec:eff}


\subsection{Perturbative Unitarity}
\label{sec:pertunit}

We consider the lowest order CP-odd operators involving the physical Higgs field,
which lead to CP-violation in conjunction of the CP-even operators in the Standard Model. A comprehensive list of operators has been presented in \cite{Gavela:2014vra}, for the purpose of this paper we limit ourselves to a few key operators, which, on the one hand, are relevant to the dimension 6 framework. On the other, we also discuss the particular example of an operator that marks the transition to phenomenologically richer simplified models. These models are just another form of EFT, in the sense that they capture key features of a UV completion at an intermediate energy scale by including additional propagating degrees of freedom. Since there are are only limited sources of CP violation in the SM, extra propagating degrees of freedom in relation with CP violation and their possible interplay with the observed Higgs phenomenology is a relevant question.

In this study we work in the broken phase of $SU(2)_L \times U(1)_Y$ and consider CP violating operators effectively up to dimension 5.
With this condition, we have the unique operator in the fermion-Higgs sector
\begin{subequations}
\begin{equation}
{\cal O}^{h f f}_4 = h \bar \psi_f \gamma_5 \psi_f \,,
\label{eq:hff_ope}	
\end{equation}
with $f$ denoting the Standard Model fermions ($f = u, d, s, c, b, t$). 

In the gauge-Higgs sector, we consider the following operators
\label{eq:cpoddoperators}
	\begin{alignat}{5}
\label{eq:cpoddoperators1}	
	{\cal O}^{h F \tilde F}_5 &= h F^{\mu \nu} \tilde F_{\mu \nu} \,, \\ 
	{\cal O}^{h h Z}_4 &=   h (\partial_\mu h)Z^\mu\,.
	\label{eq:zhh}
\end{alignat}
\end{subequations}
We use $F = (A, Z, W, G)$ as the (dual) field strengths of the photon, $Z$-boson, $W^\pm$-boson and gluon. The first class of operators is the ``standard'' set of CP violating operators that is based on a generic dimension 6 approach~\cite{Grzadkowski:2010es} and they are typical representatives of a broader class of CP-odd interactions summarised in~\cite{Gavela:2014vra}. The operators involving the dual field strength can be generated by integrating out massive fermions with CP-odd Yukawa couplings as in Eq.~\eqref{eq:hff_ope}.

The operator of Eq.~\eqref{eq:zhh} deserves a special comment as its appearance is linked to extending the dimension six EFT framework to a simplified model which is a form of EFT that contains explicit new propagating degrees of freedom. 

Let us sketch how this operator can be generated from a simplified multi-Higgs model, based on the extension of a two Higgs doublet model by a real singlet scalar. If EWSB is triggered by more than one Higgs doublets $H_j=[\phi_j^+,(v_j+h_j+is_j)/\sqrt{2}]^T$, the kinetic term  \begin{equation}
	{\cal{L}}_{\text{kin}}=|D_\mu H_1|^2+ |D_\mu H_2|^2 
    \label{eq:kin1}
\end{equation}
 leads to the massless would-be Goldstone boson to be eaten by the $Z$ boson, which is given by
 \begin{equation}
    \label{eq:goldstone}
 	s={v_1s_1 + v_2s_2\over v}
 \end{equation}
with $v=\sqrt{v_1^2+v_2^2}$ which couples 
\begin{equation}
 	{\cal{L}}_{\text{kin}}\supset m_Z \,\partial_\mu s\, Z^\mu \,.
\end{equation}
This term is removed by $R_\xi$ gauge fixing \cite{Fujikawa:1972fe}, while the CP-odd linear combination orthogonal to Eq.~\eqref{eq:goldstone}
 \begin{equation}
    \label{eq:goldstone2}
 	\tilde s={ v_2s_2 -v_2s_1\over v}
 \end{equation}
 produces the well known two-Higgs doublet interaction
 \begin{multline} 
 \label{eq:kinvt}
 {\cal{L}}_{\text{kin}} \supset {g_Z  \over 2v} \left( v_1h_2 - v_2 h_1 \right) \partial_\mu \tilde s \, Z^\mu\\ - {g_Z  \over 2v} \left( v_1 \,\partial_\mu h_2 - v_2 \,\partial_\mu h_1 \right) \tilde s \, Z^\mu\,,
 \end{multline}
which, however, cannot give rise to the operator Eq.~\eqref{eq:zhh} since the original kinetic terms Eq.~\eqref{eq:kin1} perserve CP.

If there is in addition to the two Higgs doublets a portal-type coupled real singlet scalar $S$ (which does not receive a vacuum expectation value\footnote{A non-zero $\langle S\rangle$ would not change our discussion as it only can be absorbed in a field redefinition of $H_i$ in Eq.~\eqref{eq:thdm}}.), we can postulate dimension 5 operators involving
\begin{equation}
\label{eq:thdm}
{\cal{L}}_{\text{eff}} \supset \frac{c_i}{\Lambda} S \, |D_\mu H_i|^2  
\end{equation}
(neglecting non-diagonal terms for convenience). Then we have the additional interactions in unitary gauge
\begin{equation}
\label{eq:thdm2}
	{\cal{L}}_{\text{eff}} \supset g_Z{c_2-c_1\over \Lambda } {v_1v_2 \over 2v} S \, \partial_\mu \tilde s \, Z^\mu \,.
\end{equation}
Similar terms are present for the Goldstone boson if we work in a general gauge, in the following we will, however, adopt unitary gauge for convenience. 

Explicit CP-violating terms in the two Higgs doublet potential (i.e. complex Higgs self-interactions) induce a mixing of $(h_i,\tilde{s},S)$. For instance
\begin{multline}
    V(H_1, H_2) \;\supset\; \lambda_6 |H_1|^2 (H_1^\dagger H_2) + {\text{h.c.}} \\ \;\supset \; -\, {\text{Im}(\lambda_6)}{v_1\over 2 v}\, \left( { (3 v^2-v^2_2)} \, h_1 \, \tilde s  + { v_1v_2} \, h_2 \, \tilde s  \right)  \,.
\end{multline}
If we also introduce a portal interaction
\begin{equation}	
	V(H_1,S) \; \supset \; \eta\, |H_1|^2\, S \; \supset \;\eta v_1 \, h_1 S\\
\end{equation}
we can see that mixings $\tilde s \to h_1$ and $S \to h_1$ in Eq.~\eqref{eq:thdm2} will induce Eq.~\eqref{eq:zhh}, if we understand $h_1$ as the SM-like boson in the mass basis.
Note that the CP violating interaction Eq.~\eqref{eq:zhh} cannot be introduced from the kinetic terms Eq.~\eqref{eq:kin1}, which preserves CP. 
This can explicitly be seen by the anti-symmetric structure in Eq.~\eqref{eq:kinvt}, which results in zero diagonal couplings after diagonalisation of the mass mixing matrices.
Since we are not interested in the effects of other couplings we assume that these additional states are sufficiently heavy to not immediately influence the Higgs decay phenomenology as well as unitarisation rules through additional channels opening up. We therefore assume $h$ is dominantly composed of $h_1$ in the following.\footnote{It should also be noted that the presence of multiple mixings typically yields a more SM-like phenomenology of the lightest state in terms of signal strengths as compared to minimal Higgs portal scenarios \cite{Choi:2013qra}.} Eq.~\eqref{eq:thdm} is part of the first term of a linear expansion in $S$. We will see that this particular coupling is perturbatively unconstrained and has an intriguing relation to the complex mass scheme in the SM.

In order to respect the stringent flavour constraints we only consider flavour diagonal operators
as in eq.~\eqref{eq:hff_ope} (we will comment on the impact of flavour constraints later, see also~\cite{Brod:2013cka}). With these operators we can calculate the high energy behavior of $2\to 2$ scattering amplitudes from an initial state~$i$ to a final state~$f$
\begin{multline}
	a^J_{fi}(s) = {1\over 32 \pi} \int \hbox{d}\cos\theta \, d^J_{\mu\mu'}(\cos\theta)\, {\cal{M}}_{fi}(s,\cos\theta)\\(\hbox{for $\sqrt{s}\gg m_i,m_f$})
\end{multline}
where ${\cal M}$ denotes the matrix element modulo factors of $\sqrt{2}$ for identical particles.\footnote{We choose a convention for ${\cal{M}}$ which gives rise to a real value at large $s$.} The $d^J_{\mu\mu'}$ are Wigner functions with the angular momentum $J$ that appear in the expansion of~\cite{Jacob:1959at}. $\mu,\mu'$ are calculated from the helicity differences of initial and final state of the participating particles respectively. In particular, $d^J_{00}$ is given by the Legendre polynomial as $d^J_{00} = P_J(\cos\theta)$.  Unitarity together with perturbativity (which we also have to impose for an expansion of the operators \eqref{eq:cpoddoperators} to be meaningful) requires the partial waves to be small compared to unity and critical couplings at tree level are conventionally derived from saturating 
\begin{equation}
	|a_{fi}^J(\Lambda)|=1
\end{equation}
(see Refs.~\cite{Chanowitz:1978mv,Chanowitz:1978uj} for a detailed discussion).

In the following we consider the Lagrangian
\begin{multline}
\label{eq:lagrange}
	{\cal{L}}={\cal{L}}^{\text{SM}} + C_{h h Z} \  h (\partial_\mu h)Z^\mu
	+  C_{htt} \, h \bar t \gamma^5 t \\ 	+ \sum_{F,\tilde{F}} {C_{h F \tilde F}\over v} \, {\cal O}^{h F \tilde F}_5	\,,
\end{multline}
i.e. we focus on the top quark in particular and choose the electroweak vacuum expectation value as reference scale where necessary. Typically the partial waves exhibit a hierarchy in the angular momentum $J$.

We have surveyed the list of processes relevant for unitarity violation at tree-level (see also~\cite{Lee:1977eg} for a first discussion of unitarity in the SM). These include vector, Higgs and fermion scattering, as well as combinations of the different particle species~\cite{Chanowitz:1978mv,Chanowitz:1978uj}. We find that the tightest constraints follow from the $J=0$ projections and will focus on the most constraining channels, but also mention other channels that are relevant for the discussion of the remainder of this paper. Our results are collected in Fig.~\ref{fig:j0} and we detail them below:

\begin{figure*}[!t]
	\includegraphics[width=0.65\textwidth]{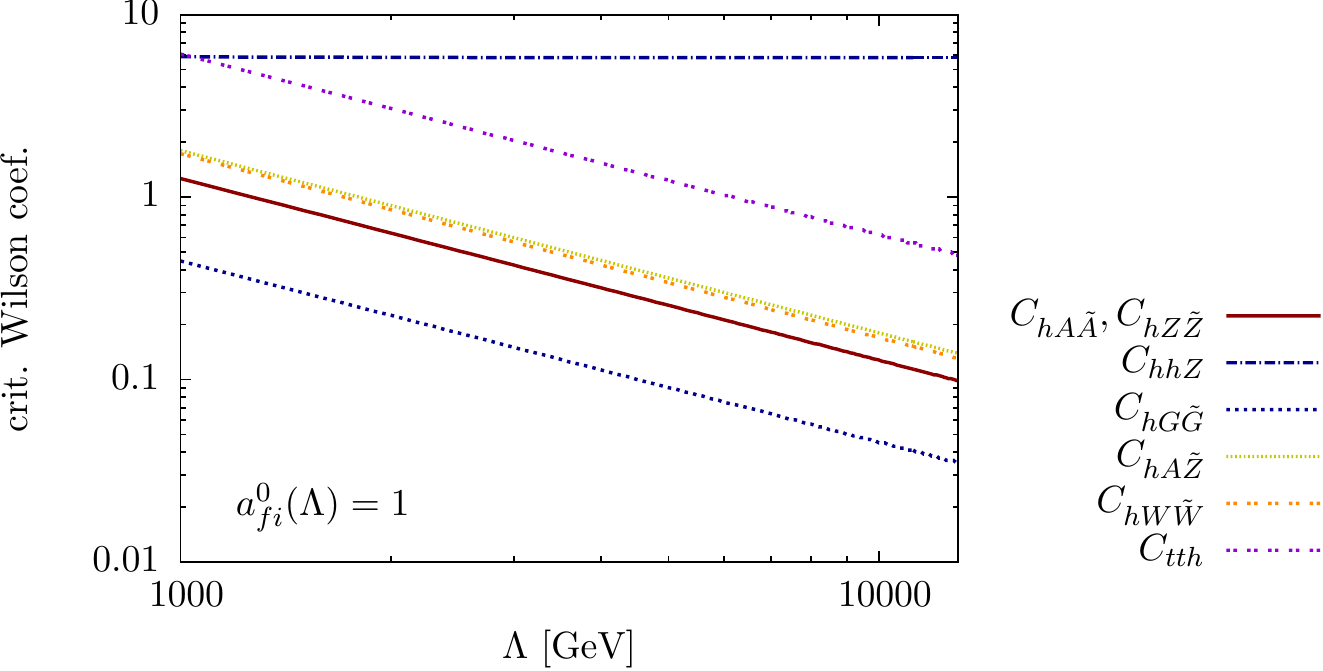}
	\caption{Selected region by the zeroth partial wave in $2\to 2$ scattering.}
	\label{fig:j0}
\end{figure*}

\begin{description}
	\item[${\cal{O}}_4^{hff}$:] We first consider fermion-fermion scattering $t\bar t\to t\bar t$, which receives contributions from the operator \eqref{eq:hff_ope}. We include the (modified) Higgs, $Z$ boson and photon intermediate states and discard the gluon contribution as it corresponds to a non-trivial color configuration, which does not interfere with the color singlet exchange. The zeroth partial wave of $t\bar t \to t\bar t$ for identical helicity (zero total angular momentum) gives rise to only weak constraints on the Wilson coefficient $|C_{htt}|\simeq 7.1$, not dependent on the scale $\Lambda$. The amplitude also quickly approaches an asymptotic value as a consequence of energy scales cancelling between the spinor normalisations and the $s$-channel suppression leading to an energy-independent value of this value. 
	
Superior bounds can be obtained from $t\bar t\to V_L V_L$. Note that due to the vertex structure induced by the operators of Eq.~\eqref{eq:cpoddoperators}, the Wilson coefficients $C_{hF\tilde F}$ do not contribute to scattering processes involving longitudinally polarized vector bosons.\footnote{
	The Feynman rule for the vertex induced by the operator \eqref{eq:cpoddoperators} is given as $\propto \epsilon_{\alpha \beta \mu \nu} p_1^\mu p_2^\nu$,
	which vanishes when contracting with the longitudinal polarization vectors of external gauge bosons: $\epsilon_{\alpha \beta \mu \nu} p_1^\mu p_2^\nu \epsilon^\alpha_L(p_1) \epsilon^\beta_L(p_2) = 0$.
} Hence, the limit obtained from inelastic fermion scattering to gauge bosons provides a way to derive stringent unitarity constraints on $C_{tth}$ without the influence of accidental cancellations between the interactions in \eqref{eq:lagrange}. It is worth noting that the CP-odd Higgs interactions therefore also exhibit a completely different unitarity-related behavior than their CP-even counterparts~\cite{Lee:1977eg}. Numerically we find that $t\bar t\to W^+_L W^-_L$ provides the most stringent constraint among these channels as the amplitude shows a $\sim C_{tth} \sqrt{s}$ behavior.
	\item[${\cal{O}}_5^{hF\tilde F}$:] We derive unitarity bounds on the Wilson coefficients $C_{hA\tilde A},C_{hZ \tilde Z}$, and $C_{hg\tilde g}$ through investigating $VV\to VV$ scattering for transverse polarisations of the participating vector bosons $V=A,Z,g$. For equal helicity and transverse $ZZ$ scattering we obtain, e.g., 
	    \begin{multline}
	    	\hspace{1cm}{\cal{M}}(Z_TZ_T\to Z_TZ_T) 
		= \\ - {4  C_{hZ\tilde Z}^2 \over  v^2}\, {s\left(s - 4m_Z^2 \right)\over s-m_h^2} + \{\text{SM}\}	 \,, 
		\end{multline}	    
		where $\{$SM$\}$ refer to the well-known results of the SM~\cite{Lee:1977eg}, which do not give rise to unitarity violation. For the massless gluons and photons we find a similar relation for the unitarity violation-driving part.
	    	    
	    The channels involving both $A$ and $Z$ introduces a cross-talk between the $C_{hA\tilde A},C_{hA\tilde Z}$ and $C_{h\tilde A Z}$ channels and the results quoted in Fig.~\ref{fig:j0} are calculated assuming $C_{hA\tilde A},C_{hA\tilde Z}=0$, which allows us to set constraints on $C_{hA\tilde Z}$ individually.
	 \item[${\cal O}^{h h Z}_4$:] $J=0$ unitarity constraints on this operator are calculated from multi-Higgs scattering.
	 For $t\bar t \to hh$ in the equal helicity case, we obtain 
	\begin{multline}
		\label{eq:tophh}
 	 \hspace{1cm}{\cal{M}} (t\bar t \to h h) = \\ {e\over 2 s_Wc_W} {m_t \sqrt{s} \over s-m_Z^2} C_{hhZ} + \{\text{SM}\} \,,
	\end{multline}
	which shows that only weak constraints can be derived from this channel as the amplitude becomes quickly negligible at energies $\sqrt{s}\gg m_Z$ even when $C_{h h Z}\neq 0$ ($c_W,s_W$ denote the cosine and sine of the Weinberg angle). This result also shows that unitarity constraints from the fermion sector are parametrically suppressed by the quark mass and that the top-quark sector will provide the most dominant unitarity constraints.
	
	$hh \to ZZ$ and $hh\to WW$ induced by ${\cal O}^{h h Z}_4$ vanish, irrespective of helicities. This also holds for $hh\to hZ_L$, leaving only $hh\to hh$ as a potentially sensitive channel to $C_{h hZ}$ for $J=0$. 
	In this channel, however, crossing symmetry guarantees that the amplitude can only have a small sensitivity on the energy of the scattering process for $s\gg m_Z^2,m_H^2$. With $s+t+u=4m_h^2$ and this cancellation only slightly affected by the different propagators of the $s,t,u$ channels for large enough energy, the unitarity constraint becomes largely insensitive to the probed energy (Fig.~\ref{fig:j0}). Amplitudes for $ZZ\to hZ$ vanish irrespective of polarisations; $WW\to hh$ does not receive contributions from ${\cal O}^{h h Z}_4$ insertions, and $hh\to hZ_L$ is suppressed by an order of magnitude compared to $hh\to hh$ at the amplitude level. 	
	  \end{description}

\begin{table}[t!]
\centering
{\renewcommand\arraystretch{1.4}
  \begin{tabular}{|c||c|c|c|} \hline
    Wilson & Most sensitive & Scaling of $|\cal M|$& limit at \\ 
    coefficient & channel & at large $s$ & $\Lambda = 5$\,TeV\\ \hline \hline    
    $C_{tth}$ & $t \bar t \to W_L^+ W_L^-$ & $C_{tth} \sqrt{s}$ &  1.24 \\
    $C_{h F \tilde F}$ & $V_T V_T \to V_T V_T$ & $C^2_{h F \tilde F} s$ & 0.26\\
    $C_{h G \tilde G}$ & $g_T g_T \to g_T g_T$ & $C^2_{h G \tilde G} s$ & 0.09\\
    $C_{h A \tilde Z}$ & $Z_T A_T \to Z_T A_T $ & $C^2_{h A \tilde Z} s$ & 0.36\\    
    $C_{hhZ}$ & $hh \to hh$ & $C_{hhZ}^2$ & 5.82\\        
    \hline
  \end{tabular}}
\caption{Representative values of perturbative unitarity constraints of the operators considered in this work at $\Lambda=5~\text{TeV}$, in addition to the most sensitive channel to unitarity constraints.}
\label{tab:unitarityconst}  
\end{table}

Table~\ref{tab:unitarityconst} summarises the constraints on the Wilson coefficients we found in this section 
based on the perturbative unitarity argument.  
Out of the operators we consider in this work, ${\cal O}^{h h  Z}_4$ is special in the sense that perturbative unitarity arguments do not limit the associated Wilson coefficient's range.
This means that a potentially large CP violation with this term could be induced by a non-perturbative or perturbative UV completion.

The operator ${\cal O}^{h f f}_4 $ has an interesting relation with the so-called complex mass scheme \cite{Nowakowski:1993iu,Papavassiliou:1996zn,Denner:1999gp,Denner:2006ic}, which continues perturbative calculations to the second Riemann sheet \cite{Passarino:2010qk} of the $S$ matrix by ``absorbing'' the Dyson-resummed imaginary part of the gauge boson two-point functions into complexification of the gauge boson masses, i.e. $m_V^2\to m_V^2 - i\Gamma_Vm_V$.\footnote{Note that introducing a constant width amounts to an ad-hoc replacement of the running width~\cite{Goria:2011wa,Passarino:2013bha}.} Through its relation with the masses, such replacements imply a complexification of the Weinberg angle as well. Typically, reordering the perturbative series, which this replacement effectively amounts to, can imply a violation of gauge invariance and therefore imply unitarity violation~(see e.g.~\cite{Kurihara:1994fz,Berends:1970bw,Baur:1995aa,Argyres:1995ym,Bhattacharyya:2012tj}). As demonstrated in \cite{Denner:1999gp,Denner:2006ic}, however, Ward and Slavnov-Taylor identities are not modified by these replacements and gauge invariance remains intact; unitarity violation can therefore not be amplified at higher energies. Allowing a complexification of the Weinberg angle, the $Z$ boson coupling to the Higgs in the SM becomes complex through its coupling $\sim e/(s_wc_w)$ where $s_w,c_w$ are sine and cosine of the Weinberg angle respectively. For a complex gauge coupling, the operator Eq.~\eqref{eq:zhh} is also generated in the SM from the Higgs kinetic term. We stress that this effect in the SM is purely spurious, but is accompanied by no unitarity limitations as a consequence of the consistency of the complex mass scheme. This also explains why we find only weak constraints on this particular operator even if it is generated in BSM scenarios.

\subsection{Unitarity sum rules for CP-odd interactions}

The $f \bar f \to V_L V_L$ processes are special in the sense that a CP-violating $\bar f f h$ coupling enters linearly, thus opening up the possibility to compensate the CP-odd operator induced unitarity violation through the appearance of an additional state. The linearity is a necessary requirement as destructive interference cannot be introduced on the quadratic coupling level. The latter appears in the SM for longitudinal gauge boson scattering as the interplay of $s,t,u$ channels and a different vertex structures. Similar ideas in the context of CP-even modifications of Higgs couplings are apparent from singlet mixing scenarios and can be generalised to the vector case giving rise to concrete phenomenological predictions, see e.g.~\cite{Birkedal:2004au,Alboteanu:2008my,Ballestrero:2011pe,Borel:2012by,Delgado:2015kxa,Englert:2015oga}. 
In concrete UV scenarios, these unitarity sum rules are always consequences of spontaneous symmetry breaking~\cite{Cornwall:1973tb,Cornwall:1974km}.

For the processes with Wilson coefficients entering at the quadratic level, unitarity cancellations of this type cannot be implemented ``by hand'' as destructive interference would require complex couplings which conflicts with the requirement of a hermitian Lagrangian. While this also applies to $f\bar f \to f\bar f$ the constraints on the Wilson coefficients from this process are only weak (Sec.~\ref{sec:pertunit}). In practice, an additional resonance that serves to cancel the growth in $\bar f f \to V_L V_L$ is not affected by unitarity considerations of $f\bar f \to f\bar f$.

\begin{table}[!t]
	\resizebox{\columnwidth}{!}{%
	{\renewcommand\arraystretch{1.4}
  \begin{tabular}{|c|c|l|} \hline
  	Vertex & Feynman rule & SM \\ 
  	\hline
  	\hline  	
  	$W^-_\alpha(p) W^+_\beta(k) A_\mu(q)$ & $- g_W^\gamma \Gamma_{\alpha, \beta, \mu}(p,k,q)$ & $g_W^\gamma = g s_W$  \\
  	$W^-_\alpha(p) W^+_\beta(k) Z_\mu(q)$ & $ g_W^Z \Gamma_{\alpha, \beta, \mu}(p,k,q)$ & $g_W^Z = g c_W$  \\  	
  	$f \bar f W_\mu^\pm$ & $g_W^f \gamma_\mu P_L$ & $g_W^f = g/2$ \\
  	$f \bar f A_\mu$ & $-g_\gamma^f \gamma_\mu$ & $g_\gamma^f = g s_W Q_f$ \\
  	$f \bar f Z_\mu$ & $\gamma_\mu (g^Z_{fL}P_L + g^Z_{fR}P_R)$ & $g^Z_{fR} = (g/c_W)(T_3^f - Q_f s_W^2)$ \\
  	                 &                                            & $g^Z_{fL} = -(g/c_W) Q_f s_W^2$ \\  	
  	                 &                                            & $g^Z_{fV} = (g^Z_{fL} + g^Z_{fL})/2$ \\  	  	                 
  	                 &                                            & $g^Z_{fA} = (g^Z_{fL} - g^Z_{fL})/2$ \\  	  	                   	                 
  	$h f \bar f$ & $-(g_h^f + i g_A^f \gamma_5)$ & $g_h^f = g m_f/(2 m_W)$ \\
  	             &                                & $g_A^f = 0$ \\
  	$h W^+_\mu W^-_\nu$ & $g_h^W g_{\mu \nu}$   & $g_h^W = g m_W$  \\  	  	             
  	$h Z_\mu Z_\nu$     & $g_h^Z g_{\mu \nu}$   & $g_h^Z = (g^2+{g'^2})^{1/2} m_Z$  \\  	  	               	
  	\hline
  \end{tabular}}
    }
  \caption{\label{tab:rules} 
  Feynman rules relevant for $f\bar f \to W^+W^-$, $P_{L,R}$ denote the right- and left-chirality projectors.}
\end{table}

We focus on $f(p_1) \bar f(p_2) \to W_L^+(q_1) W_L^-(q_2)$.  The tree-level matrix element is composed of four Feynman diagrams:
fermion-exchange $t$-channel (${\cal M}_f^t$), $\gamma/Z$-exchange $s$-channel (${\cal M}_{\gamma/Z}^s$), and $h$-exchange $s$-channel (${\cal M}_h^s$).
In the high energy limit, $\sqrt{s} \gg m_W, m_Z$ and $m_h$, these matrix elements can be written as
\begin{alignat}{5}	
{\cal M}_f^t &= - \frac{(g_W^f)^2}{m_W^2} \bar v(p_2) \Big( \slashed{q_1} P_L + \frac{m_f}{2} (1 - \gamma_5)  \Big) u(p_1) \,,
\nonumber \\
{\cal M}_\gamma^s &= \frac{g_W^\gamma g_\gamma^f}{m_W^2} \bar v(p_2) \slashed{q_1} u(p_1) \,,
\nonumber \\
{\cal M}_Z^s &= -\frac{g_W^Z}{m_W^2} 
\bar v(p_2) \Big(
\slashed{q_1} g_{fR}^Z + 2 \slashed{q_1} g^Z_{fA} P_L
- m_f g^Z_{fA} \gamma_5
\Big) u(p_1) \,,
\nonumber \\
{\cal M}_h^s &= \frac{g_W^h}{2 m_W^2} 
\bar v(p_2) \Big(
g_h^f + i g_A^f \gamma_5 
\Big) u(p_1) \,,
%
\end{alignat}
where $P_L = (1 - \gamma_5)/2$.

\begin{figure}[t!!]
	\includegraphics[width=0.45\textwidth]{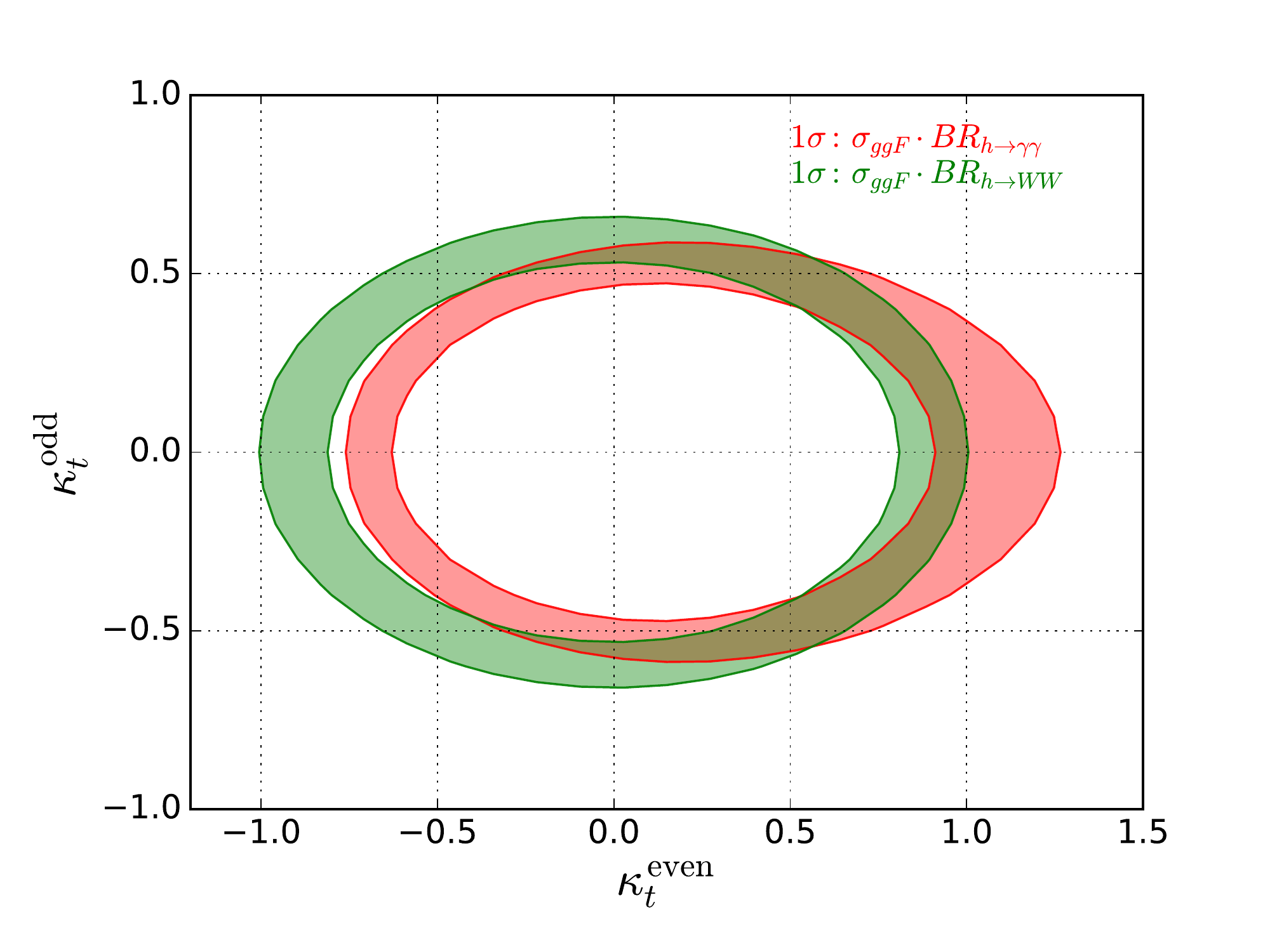}
	\caption{
	The current constraint in the $\kappa_t^{\text{even}}$ vs $\kappa_t^{\text{odd}}$ plane 
	from $\sigma_{ggF} \cdot \text{BR}(h \to \gamma \gamma)$ (red) and $\sigma_{ggF} \cdot {\text{BR}}(h \to WW)$ (green).
	The constraints from other modes are not considerable.
	}
	\label{fig:kappa_t}
\end{figure}
\begin{figure*}[!t]
	\includegraphics[width=0.45\textwidth]{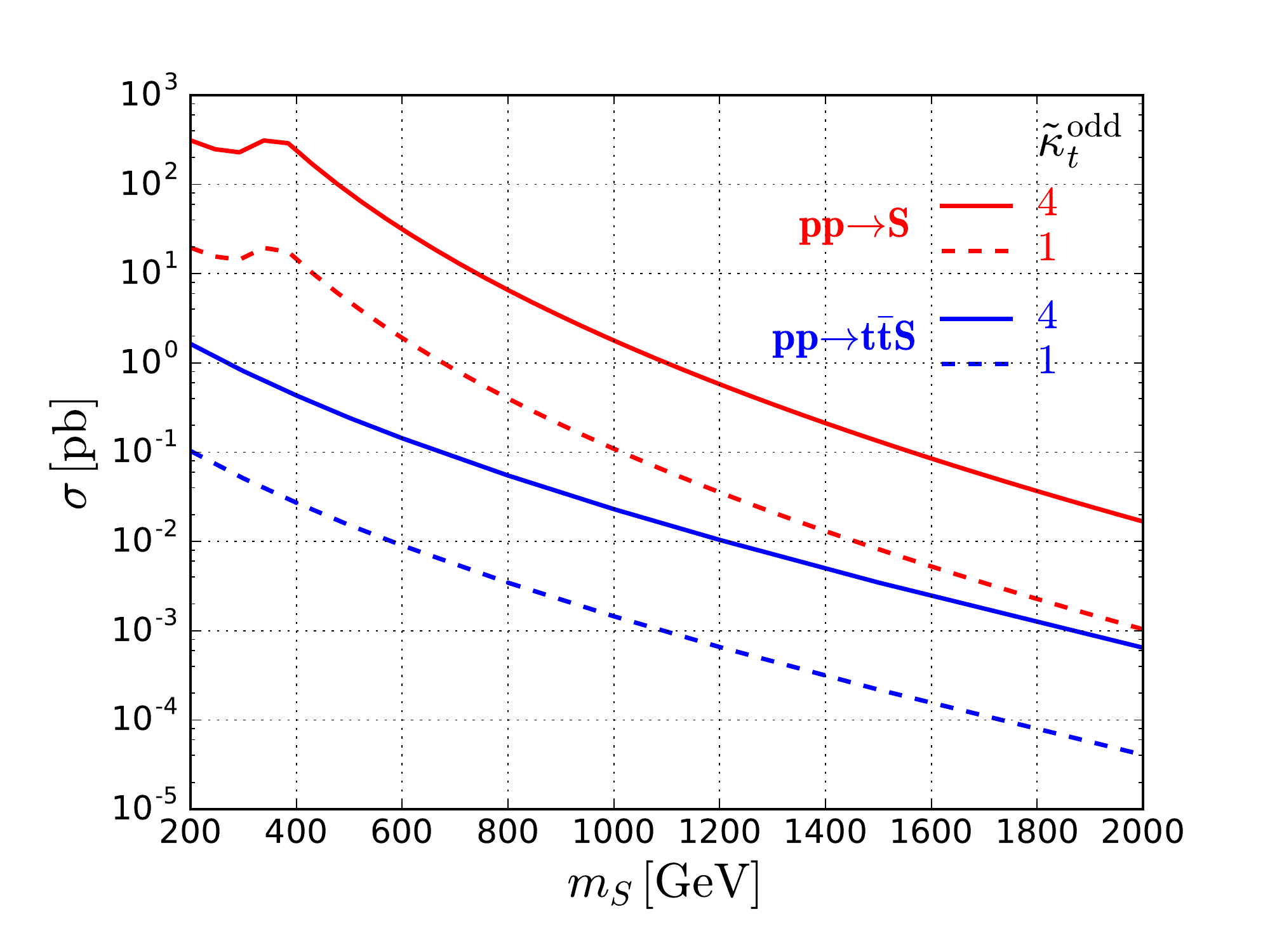}
	\hfill
	\includegraphics[width=0.45\textwidth]{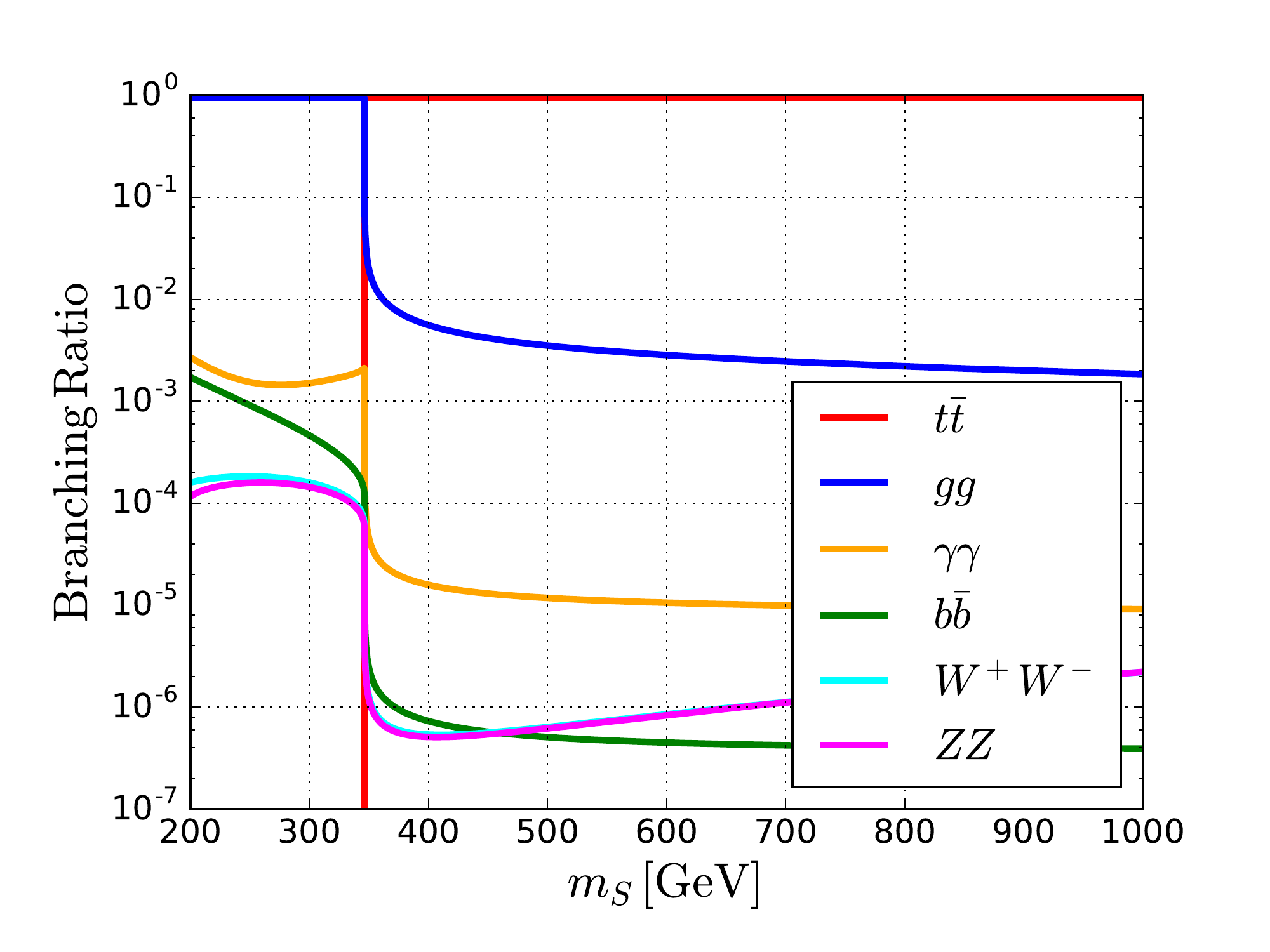}
	\caption{Cross section and branching ratios for the for the additional scalar that compensates the unitarity violation induced by ${\cal{O}}_4^{hhZ}$ for the SM-like Higgs boson for a representative choice of parameters and 13 TeV centre-of-mass energy.}
	\label{fig:hextra}
\end{figure*}
Since the fermion spinor products $v(p_2) u(p_1)$ and $\slashed{q_1}$ grow with $\sqrt{s}$,
to ensure the unitarity at high energies, the following sum rules need to be fulfilled:
\begin{eqnarray}	
(g_W^f)^2 + 2 g_W^Z g^Z_{fA} &=& 0 ~: \slashed{q_1} P_L \hspace{0.7cm} \\ 
g_W^\gamma g_\gamma^f - g_W^Z g^Z_{fR} &=& 0 ~: \slashed{q_1} \hspace{0.7cm}\\ 
(g_W^f)^2 - g_W^h g_h^f / m_f &=& 0 ~: {\bf 1}   \hspace{0.7cm}\\ 
\label{eq:crit}
(g_W^f)^2 + 2 g_W^Z g^Z_{fA} + i g_W^h g_A^f / m_f &=& 0 ~: \gamma_5 \hspace{0.7cm}
\end{eqnarray}	
In the SM, these rules are trivially satisfied, but introducing a non-zero $g_A^f$ $(=C_{tth})$ leads to a growth of the amplitude as we have seen in the previous section. However, we can mend this growth by introducing an additional scalar subject to the requirement that the imaginary part of Eq.~\eqref{eq:crit} vanishes in the high energy limit. This provides a strong constraint on the coupling of this additional scalar (the sum rules can be extended to the $f\bar f\to ZZ$ case straightforwardly). 

The currently observed Higgs coupling constraints~\cite{Khachatryan:2016vau}, dominantly from gluon fusion, can be correlated through this sum rule to arrive at a concrete prediction of how abundant this extra scalar, which we will call $S$, gets produced at the LHC and if or when we will be sensitive to such a resonant signature as a result of Higgs sector CP violation. In Fig.~\ref{fig:kappa_t}, we show the current constraints in the $\kappa_t^{\text{even}}$ vs $\kappa_t^{\text{odd}}$ plane,
where $\kappa_t^{\text{even}}$ is the relative deviation from the SM top Yukawa coupling and $\kappa_t^{\text{odd}}$ is 
the relative size of the CP-odd $t \bar t h$ coupling to the SM Yukawa coupling $(\kappa_t^{\text{odd}} \equiv C_{tth}/y_t^{\text{SM}})$.
We use the current limit $[\sigma_{ggF} \cdot {\text{BR}}(h \to \gamma \gamma)]/[\sigma_{ggF} \cdot \text{BR}(h \to \gamma \gamma)]_{\text{SM}} = 1.10^{+0.23}_{-0.22}$
and $[\sigma_{ggF} \cdot \text{BR}(h \to WW)]/[\sigma_{ggF} \cdot \text{BR}(h \to WW)]_{\text{SM}} = 0.84^{+0.17}_{-0.17}$~\cite{Khachatryan:2016vau}.
The constraints from other modes are not stronger than these two. 
Allowing the modification of the CP-even coupling $\kappa_t^{\text{even}}$, one finds the constraint $|\kappa_t^{\text{odd}}| < 0.6$.
On the other hand, if $\kappa_t^{\text{even}}$ is restricted within the range (0.8, 1.25), $|\kappa_t^{\text{odd}}| < 0.4$.\footnote{
The $\kappa_t^{\text{even}}$ and $\kappa_t^{\text{odd}}$ can be directly constrainted by the $pp \to t \bar t h$ and $pp \to t h j$ processes,
	although the sensitivity at the LHC is rather weak \cite{Ellis:2013yxa, Englert:2014pja,Kobakhidze:2014gqa,Kobakhidze:2016mfx}.
} 
The latter corresponds to the most conservative scenario, i.e. the observed Higgs state has a CP-even coupling that originates from pure SM contributions, and the CP-odd interactions originate entirely from scale separated physics that once integrated out results in an operator ${\cal{O}}_4^{hff}$. 
While a statistically significant deviation of any Higgs coupling automatically means the discovery of new physics, searches for new resonances that can be related to a potential Higgs deviation through unitarity can yield measurable effects on shorter timescales than precision Higgs physics. We therefore focus on the conservative bound on $C_{tth}$, assuming a vanishing CP-even Higgs-top modification to see if the LHC can limit the parameter space for the current constraints in the near future.

\section{Collider Phenomenology}
\label{sec:coll}

\subsection{Fermion-Higgs sector}
As a unitarity-related and potentially observable resonant effect, we show the expected gluon fusion and top-quark pair associated production cross section of the compensator state of Eq.~\eqref{eq:crit} in Fig.~\ref{fig:hextra}, as well as its branching ratios for a representative parameter choice that is in agreement with the aforementioned current Higgs measurement constraints. 
Here we assume the couplings of the SM particles to the new resonance $S$ are suppressed by factor 10 compared to the corresponding couplings to the 125 GeV Higgs boson $h$ apart from the $\bar t \gamma_5 t S$ interaction whose coupling is parametrised by $\tilde \kappa_t^{\text{odd}}$
relative to the top Yukawa coupling.
With this assumption the imaginary part of Eq.~\eqref{eq:crit} is cancelled when $\tilde \kappa_t^{\text{odd}} = 10 \kappa_t^{\text{odd}}$
because the $SWW$ coupling is 10 times smaller than that of $hWW$, and previous upper limit on $|\kappa_t^{\text{odd}}| < 0.4$
is translated to $|\tilde \kappa_t^{\text{odd}}| < 4$.
We examine $\tilde \kappa_t^{\text{odd}} = 1$ and $4$ for the production (the left panel of Fig.~\ref{fig:hextra})
and assume $\tilde \kappa_t^{\text{odd}} = 4$ for the branching ratio (the right panel of Fig.~\ref{fig:hextra}).

It becomes clear that such a state will show a dominant decay to either gluons or top quarks if the latter become kinematically accessible. This is expected as this state does not take part in the unitarisation of longitudinal gauge boson scattering. For our scenario, the production cross section is entirely dominated by CP-odd couplings, with gluon fusion being the dominant production channel.
\begin{figure}[!b]
	\centering
	\includegraphics[width=0.21\textwidth]{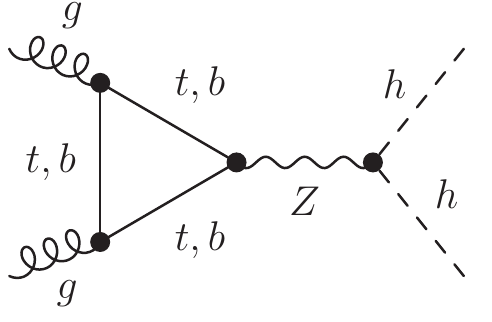}
	\caption{New contribution to Higgs pair production from gluon fusion $gg\to hh$, induced by the operator
	${\cal{O}}^{hff}_4$. We suppress the fermion flow directions as well as SM contributions.}
	\label{fig:hhfeyn}
\end{figure}

The branching fraction of $S$ to $b$-quarks $\textrm{Br}(S \to b\bar b)$ for the typical parameter point that we have chosen is even smaller than $\textrm{Br}(S\to \gamma \gamma)$. Although $B$ meson-specific triggers are available we cannot expect this process to occur at a sufficient rate to isolate it from the multi-$b$ and mistagged multi-jet backgrounds. It should be noted that due to the small Yukawa coupling of the bottom quark, unitarity constraints for $b$ channels are weak. Modifications of the $b$ sector, which we do not discuss in this work, would impact this search and provides a motivation to continue the $B$ trigger developments.

Below the $t\bar t$ threshold gives rise to an effective axion-like signature for the $S \to \gamma\gamma$ decay, see e.g.~\cite{Jaeckel:2006xm,Alekhin:2015byh}.Current constraints by ATLAS~\cite{ATLAS-CONF-2016-059} and CMS~\cite{Khachatryan:2016yec} set constraints in this mass region of about $10~\text{fb}$ in the fiducial region.

The $S\to t\bar t$ decay itself has been analysed in the context of two Higgs doublet scenarios in for example \cite{ATLAS-CONF-2016-073}. Since the top quark has a coupling structure to the CP-odd 2HDM Higgs boson $A$ of $\sim \bar t \gamma_5 t A/\tan\beta$, limits on $\tan \beta$ for specific values of $m_A$ can be related to Eq.~\eqref{eq:hff_ope}. Given the scaling with $\tan\beta$, at low values the decay of $A$ is dominantly into top pairs. In the Two Higgs doublet context the analysis of~\cite{ATLAS-CONF-2016-073} excludes a signal strength for $m_A=500~\text{GeV}$ corresponding to $\tan \beta \lesssim 1$. For a heavier mass scenario $m_A=750~\text{GeV}$ only signal strengths of $\mu \approx 3$ can be excluded, even for small values of $\tan \beta$. We can translate these constraints into $|C_{tth}(500~\text{GeV})|\lesssim 1$ and $|C_{tth}(750~\text{GeV})|\lesssim 1.7$. These values are not competitive with measurements of the SM Higgs on-shell signal strength. However, the analysis of \cite{ATLAS-CONF-2016-073} clearly shows that we can expect a tremendous improvement with more searches and statistics.


\begin{figure}[!b]
	\includegraphics[width=0.43\textwidth]{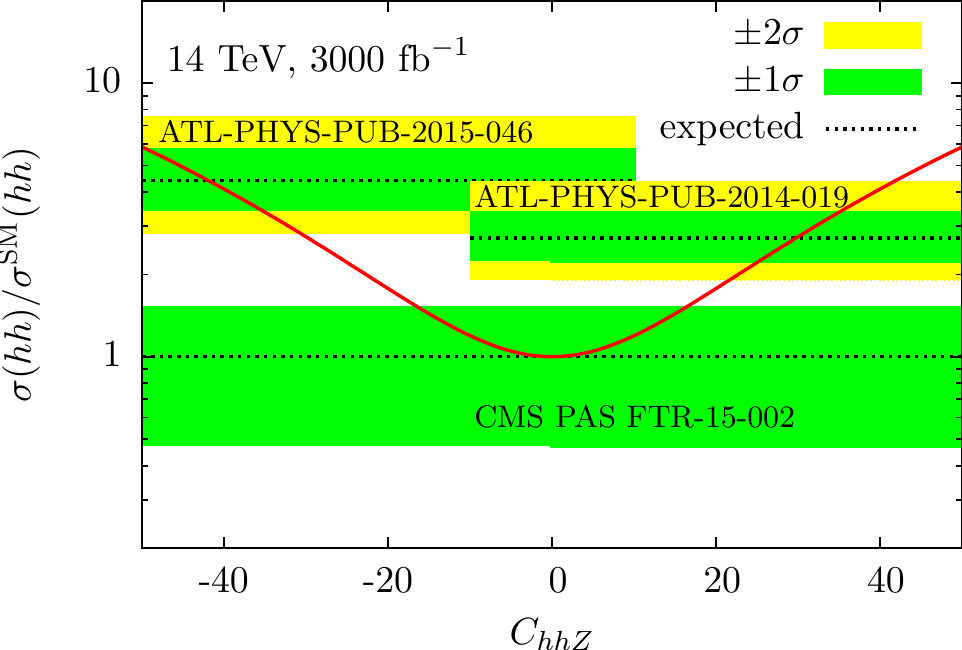}
	\caption{The expected exclusions of ATLAS~\cite{ATL-PHYS-PUB-2014-019,ATL-PHYS-PUB-2015-046} and CMS~\cite{CMS-PAS-FTR-15-002} for the high-luminosity (HL)-LHC (3000 fb$^{-1}$, 14 TeV) for $pp \to hh+X$, overlaid by the di-Higgs cross section as a function of $C_{hhZ}$ relative to the SM expectation. To highlight the different ATLAS exclusions, we do not plot them across the entire Wilson coefficient range.}
	\label{fig:hh}
\end{figure}

\begin{figure}[!t]
	\centering
	\includegraphics[width=0.2\textwidth]{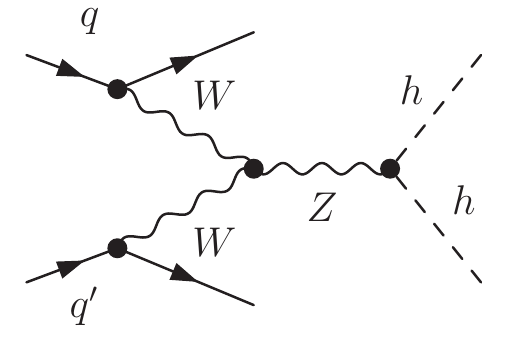}
	\caption{Representative Feynman diagram contributing to Higgs pair production from weak boson fusion $pp\to hhjj$, induced be the operator ${\cal{O}}^{hhZ}_4$.}
	\label{fig:wbfhh}
\end{figure}
\begin{figure}[!t]
	\centering
	\includegraphics[width=0.44\textwidth]{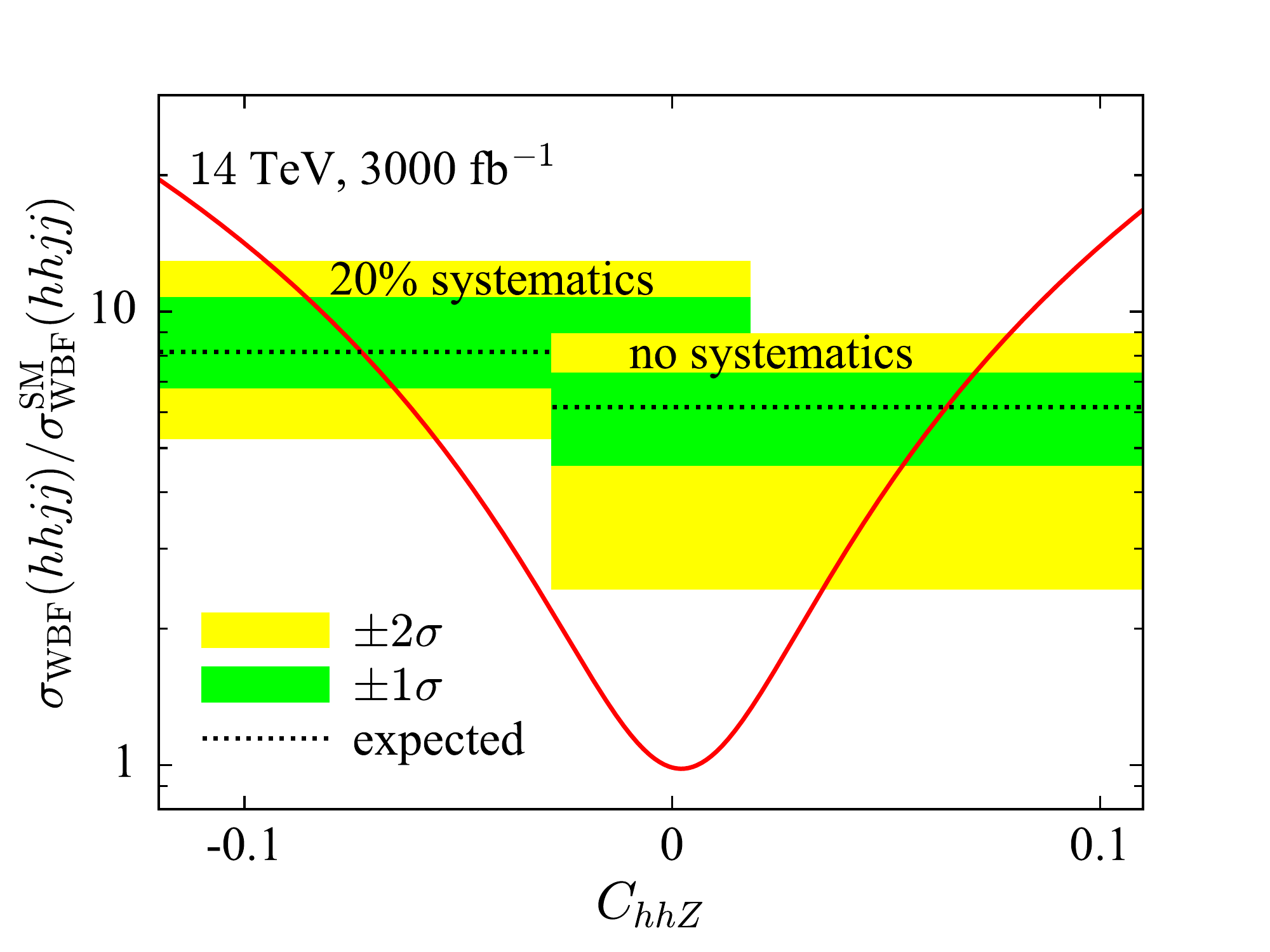}
	\caption{The expected exclusion for the high-luminosity (HL)-LHC (3000 fb$^{-1}$, 14 TeV) for the WBF-induced component of $pp \to hhjj$ using the analysis detailed in the text, overlaid by the cross section as a function of $C_{hhZ}$ relative to the SM expectation. Two different systematics scenarios are assumed.}
	\label{fig:hhjj}
\end{figure}

\subsection{Gauge-Higgs sector}
Turning to CP violation in the gauge Higgs sector, we focus on the operator ${\cal O}^{h h Z}_4$, where unitarity constraints are weak.\footnote{The physics of $hF\tilde F$ operators has been discussed in detail in the literature (e.g.~\cite{Gao:2010qx,Bolognesi:2012mm,Englert:2012xt,Ellis:2012xd,Choi:2012yg,Maltoni:2013sma,Artoisenet:2013puc}) and we will not discuss it in detail.} Such an operator will impact multi-Higgs final states (see~\cite{Baur:2002rb,Baur:2003gp,Dolan:2012rv,Papaefstathiou:2012qe,deLima:2014dta}. The dominant process of this type is Higgs pair production through gluon fusion $gg\to hh$, which can receive a new $Z$ boson-mediated contribution from ${\cal O}^{h h Z}_4$.\footnote{Note that the bottom contribution needs to be included to avoid spurious loop singularities related to $SU(3)^2\times SU(2)$ anomaly cancellations.} However, since $t\bar t \to hh$ does not give rise to an energy-dependent unitarity constraint, we can already anticipate that the absorptive parts of the $gg \to hh$ will be largely unaffected. 

Both ATLAS and CMS have published results on extrapolated sensitivity yields of Higgs pair production at the LHC~\cite{ATL-PHYS-PUB-2015-046,ATL-PHYS-PUB-2014-019,CMS-PAS-FTR-15-002}, using or even combining the $hh\to b\bar b \gamma\gamma$ and $hh\to b\bar b \tau^+\tau^-$ channels \cite{Baur:2002qd,Baur:2003gp,Dolan:2012rv}. In Fig.~\ref{fig:hh} we compare these extrapolations with the expected enhancement of $pp\to hh$ at the LHC due to the operator ${\cal{O}}^{hhZ}_4$. The most optimistic constraints that can be set from this channel result from the 1.9-$\sigma$ significance reported by CMS for the $b\bar b \gamma\gamma + b\bar b \tau^+\tau^-$ combination \cite{CMS-PAS-FTR-15-002}, which translates into a Wilson coefficient constraint
\begin{equation}
	|C_{hhZ}|\lesssim 16.5>4\pi\,.
\end{equation}
This constraint is weaker than the perturbative constraint, namely it does not play any role in the regime where
Eq.~\eqref{eq:lagrange} can be understood as a trustable series expansion.
Also, this constraint does not probe the unitarity limit imposed by $hh\to hh$ scattering, see Fig.~\ref{fig:j0}.
Although this result is expected in the light of our unitarity discussion of Sec.~\ref{sec:pertunit}, we are forced to draw somewhat unfortunate conclusion that the most dominant (and hence best motivated) di-Higgs production mechanism is unlikely to improve constraints beyond the theoretical bounds within the remit of perturbation theory.

The observed sensitivity of $pp\to hh$ to ${\cal O}^{h h Z}_4$ dominantly arises contributions of Eq.~\eqref{eq:tophh}, which contribute to the imaginary part of loop-induced $gg\to hh$ near the the threshold $m(hh)\simeq 2m_t$ through modifying the interference pattern that exists in gluon fusion between the box- and triangle-induced amplitude contributions. While there is an interference between the SM triangle- and box-contributions (identical to the off-shell interference in $pp\to ZZ$ \cite{Kauer:2012hd}), the modifications induced by ${\cal{O}}^{hhZ}_4$ quickly die out for larger di-Higgs invariant masses. This kinematic suppression cannot be circumvented, but this line of thought points to a different channel that accesses a distinct kinematic configuration of ${\cal{O}}^{hhZ}_4$, which is not probed by the unitarity constraints of Fig.~\ref{fig:j0} - di-Higgs production through weak boson fusion (WBF) \cite{Baglio:2012np,Dolan:2013rja,Frederix:2014hta,Dolan:2015zja}, which accesses $t$-channel virtual massive gauge bosons, Fig.~\ref{fig:wbfhh}.

This leads to a sizeable contribution to WBF-induced di-Higgs production, which can be investigated through the $hhjj$ final state using the full HL-LHC dataset \cite{Baglio:2012np,Dolan:2013rja,Frederix:2014hta,Dolan:2015zja,Bishara:2016kjn}. 
We make a projection of the HL-LHC's expected sensitivity to $C_{h h Z}$ by generating hadron-level $hhjj$ events using \textsc{MadEvent} \cite{Alwall:2014hca} and \textsc{Herwig}~\cite{herwig}. Following \cite{Dolan:2015zja} we focus on the $hh \to \tau^+\tau^- b \bar{b}$ final state and simulate $t\bar{t}jj$, $t\bar{t}h$, $Zhjj$, $ZZjj$, and $ZWWjj$ as backgrounds. We impose the following selections in order to improve the sensitivity and suppress the QCD-mediated signal component, which is insensitive to $C_{h h Z}$ and therefore acts as an additional irreducible background to the analysis:

\begin{enumerate}
\item We simulate a staggered two tau-trigger by requiring two taus with $p_T \ge 29,20~\text{GeV}$ in $|\eta_\tau|<2.5$ and further apply a flat tau tagging efficiency of 70\%. 
  We define jets by finding $R = 0.4$ anti-$k_T$ objects using \textsc{FastJet}~\cite{fastjet} and requiring $p_{T,j} \ge 25$ GeV and $| \eta_j | \le 4.5$.
\item We $b$-tag the two hardest jets with an efficiency of 70\%
  and fake rate of 1\% within $|\eta_j|<2.5$. If either of these 
  overlaps with a tau we veto the event. We require at least two 
  additional jets which are referred to as tagging jets and refer
  to the two leading ones as $j_1$ and $j_2$.
\item We require  $|m_{bb} - m_h| < 15$ GeV, $|m_{\tau \tau} - m_h| < 25$ GeV,
      and $m_{bb\tau \tau} > 400$ GeV.
\item Finally we require the two leading tagging jets to be widely separated
      in $\eta$, such that $\Delta \eta(j_1,j_2) \ge 5$.
\end{enumerate}

Since no analysis of the arguably complicated $pp\to hhjj$ process has been performed by the experiments yet, we proceed to compute expected cross section limits following \cite{Dolan:2015zja} to estimate the limits that can be set on $C_{h h Z}$ using the CLs method \cite{cls1,cls2}. The signal and background cross sections after all selections are applied are given in Table.\ref{tab:xs}. To show the impact of uncertainties we provide limits based on using 20\% flat background systematics as well as excluding systematics for comparison (Fig.~\ref{fig:hhjj}).

As can be seen, accessing the $t$-channel $W$ and $Z$ bosons in the initial state enhance the sensitivity to $C_{h h Z}$ way below the unitarity limit, with expected constraints
\begin{equation}
	|C_{h h Z}| \lesssim 0.06\,,
\end{equation}
within the validity of the perturbative expansion of Eq.~\eqref{eq:lagrange}.

\begin{table}[!t]
\renewcommand\arraystretch{1.3}
    \begin{tabular}{lr}
    \toprule
    Sample      	&  After Selection [fb] \\
    \botrule
    $hhjj$ (WBF)	& \hspace{0.1cm}  \num{0.001485}  \\
    \hline
    $hhjj$ (GF)		& \hspace{0.1cm}  \num{0.0005378} \\
    $t\bar{t}jj$ 	& \hspace{0.1cm}  \num{0.01801}	\\
    $t\bar{t}h$		& \hspace{0.1cm}  \num{0.00005658}\\
    $Zhjj$		& \hspace{0.1cm}  \num{0.0001026}\\
    $ZZjj$		& \hspace{0.1cm}  \num{0.0000007639}	\\
    $ZWWjj$		& \hspace{0.1cm}  \num{0.0000002039}  \\
    \hline
    total background	& \hspace{0.1cm}  \num{0.01870}	\\
    \botrule    
    $S/B$		& \hspace{0.3cm}  1/12.60    \\
 \botrule
     \end{tabular}
     \caption{Cross sections for the signal and backgrounds (including the gluon fusion-produced $hhjj$ component) in the $hhjj$ analysis after all cuts have been applied. 
     The signal cross section shown here is calculated assuming the Standard Model hypothesis.
     Turning on $C_{h h Z}$ induces a large contribution to the signal yield which makes limit-setting possible despite the low $S/B$ in the Standard Model case.
     }
 \label{tab:xs}
\end{table}

\section{Summary And Conclusions}
\label{sec:conclusions}
After the Higgs discovery the search for its particular role in the mechanism of electroweak symmetry breaking is underway. The high energy physics community has moved to largely model-independent strategies based on the application of effective field theory techniques. Unitarity as well as perturbativity set important constraints on using these tools in searches for new phenomena beyond the well-established SM. If nature indeed chooses a new physics scale well separated from the electroweak scale, then the natural question we need to address at this stage of the LHC programme is how large effective interactions can become. While this is a largely model-dependent question, perturbative unitarity provides an important guideline in reflecting hierarchies among Wilson coefficients at the $\sim\text{TeV}$ measurement scale that can be excited by a consistent (perturbative) UV completion. This question becomes particularly interesting when we turn to CP violation in the Higgs sector as most interactions induce unitarity violation even for relatively small Wilson coefficient choices. In some cases this unitarity violation can be mended through a new degree of freedom, whose interactions are governed by probability conservation, which gives rise to concrete phenomenological implications.

Considering unitarity arguments for a number of effective CP-violating interactions in the Higgs sector, we have identified two phenomenologically relevant directions: The search for a new Higgs-like state, which compensates CP-violating interactions of the Higgs with top quarks, and searches for enhancements of multi-Higgs production from weak boson fusion due to CP-violating interactions which are largely unconstrained by unitarity considerations.


\medskip

\noindent{\it{Acknowledgements}} --- We thank Apostolos Pilaftsis for helpful conversations. KS thanks Richard Ruiz for useful discussion.  KN is supported in part by the University of Glasgow College of Science \& Engineering through a PhD scholarship. MS is supported in part by the European Commission through the ``HiggsTools'' Initial Training Network PITN-GA-2012-316704.
The work of KS is partially supported by the National Science Centre, Poland, under research grants
DEC-2014/15/B/ST2/02157 and DEC-2015/18/M/ST2/00054.


\bibliography{references}

\end{document}